\documentclass[brazil,english,showkeys, superscriptaddress, nofootinbib]{revtex4}
\usepackage[T1]{fontenc}
\usepackage[latin9]{inputenc}
\usepackage{color}
\usepackage{babel}
\usepackage{float}
\usepackage{amsmath}
\usepackage{amssymb}
\usepackage{graphicx}
\usepackage[unicode=true,pdfusetitle,
 bookmarks=true,bookmarksnumbered=false,bookmarksopen=false,
 breaklinks=true,pdfborder={0 0 0},backref=false,colorlinks=true]
 {hyperref}
\usepackage{breakurl}

\makeatletter

\providecommand{\tabularnewline}{\\}

\@ifundefined{textcolor}{}
{%
 \definecolor{BLACK}{gray}{0}
 \definecolor{WHITE}{gray}{1}
 \definecolor{RED}{rgb}{1,0,0}
 \definecolor{GREEN}{rgb}{0,1,0}
 \definecolor{BLUE}{rgb}{0,0,1}
 \definecolor{CYAN}{cmyk}{1,0,0,0}
 \definecolor{MAGENTA}{cmyk}{0,1,0,0}
 \definecolor{YELLOW}{cmyk}{0,0,1,0}
}

\hypersetup{colorlinks=true,citecolor=blue,linkcolor= blue,urlcolor= cyan,filecolor= green}

\makeatother

\begin{document}

\title{Computing partial traces and reduced density matrices}

\author{Jonas Maziero}

\email{jonas.maziero@ufsm.br}

\address{Departamento de F\'isica, Centro de Ci\^encias Naturais e Exatas, Universidade Federal de Santa Maria, Avenida Roraima 1000, 97105-900, Santa Maria, RS, Brazil}

\address{Instituto de F\'isica, Facultad de Ingenier\'ia, Universidad de la Rep\'ublica, J. Herrera y Reissig 565, 11300, Montevideo, Uruguay}
\begin{abstract}
Taking partial traces for computing reduced density matrices, or related
functions, is a ubiquitous procedure in the quantum mechanics of composite
systems. In this article, we present a thorough description of this
function and analyze the number of elementary operations (ops) needed,
under some possible alternative implementations, to compute it on
a classical computer. As we notice, it is worthwhile doing some analytical
developments in order to avoid making null multiplications and sums,
what can considerably reduce the ops. For instance, for a bipartite
system $\mathcal{H}_{a}\otimes\mathcal{H}_{b}$ with dimensions $d_{a}=\dim\mathcal{H}_{a}$
and $d_{b}=\dim\mathcal{H}_{b}$ and for $d_{a},d_{b}\gg1$, while
a direct use of partial trace definition applied to $\mathcal{H}_{b}$
requires $\mathcal{O}(d_{a}^{6}d_{b}^{6})$ ops, its optimized implementation
entails $\mathcal{O}(d_{a}^{2}d_{b})$ ops. In the sequence, we regard
the computation of partial traces for general multipartite systems
and describe Fortran code provided to implement it numerically. We
also consider the calculation of reduced density matrices via Bloch's
parametrization with generalized Gell Mann's matrices.
\end{abstract}

\keywords{quantum mechanics, composite systems, partial trace, reduced density
matrix, Bloch parametrization, Gell Mann matrices}

\pacs{03.65.-w, 03.67.-a, 03.65.Yz}

\maketitle

\section{Introduction}

When calculating certain functions of quantum systems, in many instances
the running time of classical computers increases exponentially with
the number of elementary parts that compose those systems. This issue
is a hurdle to current research in many areas of science. But it is
also a motive for the quest towards the construction of a large scale
quantum computer \cite{Feynman82,Loyd,Ito}. For now, we have to resort
to several other alternative techniques, with which one can extract
approximate information about quantum systems using only the available
classical computing power.

Among those methods, some famous examples are: stochastic Monte Carlo
simulations \cite{Krauth,Binder,Kroese,Lutsyshyn}, mean field approximations
\cite{Kadanoff,Liu_mft,Sen,Yamamoto}, density functional theory \cite{Capelle,Ullrich,Jones,Kanjouri_DFT},
renormalization group \cite{Fisher_RG,Kadanoff_RG,Alvarez,Osborne_RG},
and matrix product states and projected entangled pair states \cite{Orus,Cirac,Troyer,Karrasch}.
On the other hand, recently several authors have shown that some general
patterns of the many-body behavior, usually accessed in the thermodynamical
limit, may be disclosed by analyzing systems with a moderate number
of particles \cite{Campbell0,Monroe,Giorgi,Campbell1,Kwek,Liu,Sen_SwNqb,Campbell2}.
In such kind of investigation, it is desirable to use a system as
large as practically possible. And for that purpose we would like
to optimize the implementation of basic and frequently used functions
in order to reduce the computation time as much as possible.

In the quantum mechanics of composite systems, one ubiquitous function
is the partial trace (PTr) \cite{Nielsen=000026Chuang,Preskill,Paris,Wilde}.
The PTr function has a unique place, for instance, for the computation
of reduced density matrices and related functions. In spite of the
physical interpretation of the partial trace not being a trivial matter
\cite{Garola,Fortin}, it's mathematical and operational meaning is
well established \cite{Nielsen=000026Chuang,Preskill,Paris,Wilde}.
Besides, the PTr appears very frequently, for instance, in the context
of correlation quantifiers (mutual information \cite{Winter,Maziero_MI},
quantum entanglement \cite{Bennett_EE,Okano,Fritzche,Loss}, quantum
discord \cite{Maziero_RevD,Vedral_rd,Maziero_Opt}, etc), in the generation
of random density matrices \cite{Maziero_rrho,Maziero_Fort}, and
in investigations regarding phase transitions \cite{Sacramento,Zeng}.
It is also a fundamental ingredient in the quantum marginal and extension
problems \cite{Bellomo2,Christandl,Modi,Viola1,Viola2,Chen1}, for
the strong subadditivity property of von Neumann entropy and related
results \cite{Bathia,Petz_SSAeq,Lieb_ESSA}, and in the theories of
quantum measurement and decoherence \cite{Schlosshauer,Zurek_Dar,Maziero_Zimmer,Fritzsche2}.

Our aim here is to examine in details the partial trace function with
special focus on its numerical calculation. The remainder of the article
is structured as follows. In Sec. \ref{PTr_def} we present two definitions
for the PTr involving bipartitions of a system, verify their equivalence,
and discuss the uniqueness of the partial trace function. In Sec.
\ref{num_bip} we address the numerical calculation of the partial
trace, firstly via its direct implementation (Sec. \ref{sec:1st})
and afterwards using two levels of optimization which are obtained
simply by avoiding making null multiplications and sums (Secs. \ref{sec:2nd}
and \ref{sec:3rd}). In Sec. \ref{PTr_mult}, the calculation of the
PTr for the general case of multipartite systems is regarded; and
Fortran code produced to implement the PTr numerically is described.
We consider the computation of the partial trace via Bloch's parametrization
with generalized Gell Mann's matrices in Sec. \ref{sec:su}. A brief
summary of the article is included in Sec. \ref{sec:cc}.

\section{Partial traces for bi-partitions}

\label{PTr_def}

Let $O$ be a linear operator defined in the Hilbert space $\mathcal{H}$,
that is to say $O:\mathcal{H}\rightarrow\mathcal{H}$. The space composed
by these operators is denoted by $\mathcal{L}(\mathcal{H})$. As a
prelude, let us recall that the trace of $O$ is a map $\mathrm{Tr}:\mathcal{L}(\mathcal{H})\rightarrow\mathbb{C}$
defined as the sum of the diagonal elements of $O$ when it is represented
in a certain basis $|\psi_{j}\rangle\in\mathcal{H}$, i.e., $\mathrm{Tr}(O)=\sum_{j=1}^{d}\langle\psi_{j}|O|\psi_{j}\rangle,$
with $d$ being the dimension of $\mathcal{H}$ \cite{Arfken}.

By its turn, in the quantum mechanics of composite systems with Hilbert
space $\mathcal{H}=\mathcal{H}_{a}\otimes\mathcal{H}_{b}$, the \emph{partial
trace function}, taken over sub-system $b$, can be defined as \cite{Watrous}
\begin{equation}
\mathrm{Tr}_{b}(O)=\sum_{j=1}^{d_{b}}(\mathbb{I}_{a}\otimes\langle b_{j}|)O(\mathbb{I}_{a}\otimes|b_{j}\rangle),\label{eq:ptr1}
\end{equation}
with
\begin{equation}
|b_{j}\rangle=[b_{j1}\mbox{ }b_{j2}\mbox{ }\cdots\mbox{ }b_{jd_{b}}]^{t}
\end{equation}
being any orthonormal basis for $\mathcal{H}_{b}$, $\langle b_{j}|=|b_{j}\rangle^{\dagger}$,
$d_{b}=\dim\mathcal{H}_{b}$, and $\mathbb{I}_{b}$ is the identity
operator in $\mathcal{H}_{b}$ ($X^{t}$ denotes the transpose of
$X$ and $X^{\dagger}$ stands for its conjugate transpose). So the
partial trace is a map
\begin{equation}
\mathrm{Tr}_{b}:\mathcal{L}(\mathcal{H})\rightarrow\mathcal{L}(\mathcal{H}_{a});
\end{equation}
and the analogous definition follows for $\mathrm{Tr}_{a}:\mathcal{L}(\mathcal{H})\rightarrow\mathcal{L}(\mathcal{H}_{b})$.

It is worthwhile observing here that the definition above is equivalent
to \emph{another definition} which appears frequently in the literature
\cite{Nielsen=000026Chuang}:
\begin{equation}
\mathrm{Tr}_{b}(|a\rangle\langle a'|\otimes|b\rangle\langle b'|)=|a\rangle\langle a'|\otimes\mathrm{Tr}_{b}(|b\rangle\langle b'|).\label{eq:ptr_def2}
\end{equation}
In the last equation $|a\rangle,|a'\rangle\in\mathcal{H}_{a}$ and
$|b\rangle,|b'\rangle\in\mathcal{H}_{b}$ are generic vectors in the
corresponding Hilbert spaces. In order to verify this assertion, let
us use two basis $|a_{j}\rangle\in\mathcal{H}_{a}$ and $|b_{k}\rangle\in\mathcal{H}_{b}$
and the related completeness relations to write 
\begin{equation}
O=(\mathbb{I}_{a}\otimes\mathbb{I}_{b})O(\mathbb{I}_{a}\otimes\mathbb{I}_{b})=\sum_{j,l=1}^{d_{a}}\sum_{k,m=1}^{d_{b}}(\langle a_{j}|\otimes\langle b_{k}|O|a_{l}\rangle\otimes|b_{m}\rangle)|a_{j}\rangle\langle a_{l}|\otimes|b_{k}\rangle\langle b_{m}|.\label{eq:Orep}
\end{equation}
The linearity of the partial trace and $\mathrm{Tr}_{b}(|b_{k}\rangle\langle b_{m}|)=\sum_{l=1}^{d_{b}}\langle b_{l}|b_{k}\rangle\langle b_{m}|b_{l}\rangle=\delta_{lk}\delta_{ml}$
(we applied the base independence of the trace function) lead to 
\begin{equation}
\mathrm{Tr}_{b}(O)=\sum_{j,l=1}^{d_{a}}\sum_{k=1}^{d_{b}}|a_{j}\rangle(\langle a_{j}|\otimes\langle b_{k}|)O(|a_{l}\rangle\otimes|b_{k}\rangle)\langle a_{l}|=\sum_{k=1}^{d_{b}}(\sum_{j=1}^{d_{a}}|a_{j}\rangle\langle a_{j}|)\otimes\langle b_{k}|O(\sum_{l=1}^{d_{a}}|a_{l}\rangle\langle a_{l}|)\otimes|b_{k}\rangle,\label{eq:ptr_def2c}
\end{equation}
which is equivalent to the definition in Eq. (\ref{eq:ptr1}). To
obtain the last equality in Eq. (\ref{eq:ptr_def2c}), we verified
that
\begin{equation}
(|a\rangle\otimes|b\rangle)\langle a'|=|a\rangle\langle a'|\otimes|b\rangle
\end{equation}
for any vectors $|a\rangle,|a'\rangle\in\mathcal{H}_{a}$ and $|b\rangle\in\mathcal{H}_{b}$.

Another important fact about the the partial trace function $\mathrm{Tr}_{b}(O)$
is that it is the only function $f:\mathcal{L}(\mathcal{H}_{a}\otimes\mathcal{H}_{b})\rightarrow\mathcal{L}(\mathcal{H}_{a})$
such that $\mathrm{Tr}_{ab}(A\otimes\mathbb{I}_{b}O)=\mathrm{Tr}_{a}(Af(O))$,
for generic linear operators $A\in\mathcal{L}(\mathcal{H}_{a})$ and
$O\in\mathcal{L}(\mathcal{H}_{a}\otimes\mathcal{H}_{b})$. To prove
this assertion, let us start \emph{assuming} that $f(O)=\mathrm{Tr}_{b}(O)$.
Then, using $O$ as written in Eq. (\ref{eq:Orep}) leads to
\begin{equation}
\mathrm{Tr}_{b}(O)=\sum_{k,l,m,n}(\langle a_{k}|\otimes\langle b_{l}|O|a_{m}\rangle\otimes|b_{n}\rangle)|a_{k}\rangle\langle a_{m}|\otimes\delta_{nl}=\sum_{k,l,m}(\langle a_{k}|\otimes\langle b_{l}|O|a_{m}\rangle\otimes|b_{l}\rangle)|a_{k}\rangle\langle a_{m}|.
\end{equation}
Now, utilizing the eigen-decomposition $A=\sum_{j}a_{j}|a_{j}\rangle\langle a_{j}|$
we shall have 
\begin{eqnarray}
\mathrm{Tr}_{a}(Af(O)) & = & \mathrm{Tr}_{a}(A\mathrm{Tr}_{b}(O))\\
 & = & \mathrm{Tr}_{a}(\sum_{j}a_{j}|a_{j}\rangle\langle a_{j}|\sum_{k,l,m}\langle a_{k}|\otimes\langle b_{l}|O|a_{m}\rangle\otimes|b_{l}\rangle|a_{k}\rangle\langle a_{m}|)\\
 & = & \sum_{j}a_{j}\sum_{k,l,m}(\langle a_{k}|\otimes\langle b_{l}|O|a_{m}\rangle\otimes|b_{l}\rangle)\underset{=\delta_{jm}\delta_{jk}}{\underbrace{\mathrm{Tr}_{a}(|a_{j}\rangle\langle a_{j}|a_{k}\rangle\langle a_{m}|)}}\\
 & = & \sum_{j,l}\langle a_{j}|\otimes\langle b_{l}|a_{j}\mathbb{I}_{a}\otimes\mathbb{I}_{b}O|a_{j}\rangle\otimes|b_{l}\rangle=\sum_{j,l}\langle a_{j}|\otimes\langle b_{l}|a_{j}\sum_{k}|a_{k}\rangle\langle a_{k}|\otimes\mathbb{I}_{b}O|a_{j}\rangle\otimes|b_{l}\rangle\\
 & = & \sum_{j,l}\langle a_{j}|\otimes\langle b_{l}|\sum_{k}a_{k}|a_{k}\rangle\langle a_{k}|\otimes\mathbb{I}_{b}O|a_{j}\rangle\otimes|b_{l}\rangle=\sum_{j,l}\langle a_{j}|\otimes\langle b_{l}|A\otimes\mathbb{I}_{b}O|a_{j}\rangle\otimes|b_{l}\rangle\\
 & = & \mathrm{Tr}_{ab}(A\otimes\mathbb{I}_{b}O).
\end{eqnarray}

To complete the proof we \emph{assume} that $\mathrm{Tr}_{a}(Af(O))=\mathrm{Tr}_{ab}(A\otimes\mathbb{I}_{b}O)$
and use a basis of linear operators $\Upsilon_{j}\in\mathcal{L}(\mathcal{H}_{a})$
to write \cite{Paris}
\begin{equation}
f(O)=\sum_{j=1}^{d_{a}^{2}}\mathrm{Tr}_{a}(\Upsilon_{j}^{\dagger}f(O))\Upsilon_{j}=\sum_{j=1}^{d_{a}^{2}}\mathrm{Tr}_{ab}(\Upsilon_{j}^{\dagger}\otimes\mathbb{I}_{b}O)\Upsilon_{j}=\sum_{j=1}^{d_{a}^{2}}\mathrm{Tr}_{a}(\Upsilon_{j}^{\dagger}\mathrm{Tr}_{b}(O))\Upsilon_{j}.
\end{equation}
Above, the second equality is obtained applying $\mathrm{Tr}_{a}(Af(O))=\mathrm{Tr}_{ab}(A\otimes\mathbb{I}_{b}O)$
and the last equivalence follows from $\mathrm{Tr}_{a}(A\mathrm{Tr}_{b}(O))=\mathrm{Tr}_{ab}(A\otimes\mathbb{I}_{b}O)$,
with $A=\Upsilon_{j}^{\dagger}$. Hence the uniqueness of the decomposition
of an element of a Hilbert space in a given basis \cite{Arfken} implies
in the uniqueness of the partial trace function, i.e., $f(O)\equiv\mathrm{Tr}_{b}(O)$.
Summing up, we proved that 
\begin{equation}
f(O)=\mathrm{Tr}_{b}(O)\Longleftrightarrow\mathrm{Tr}_{ab}(A\otimes\mathbb{I}_{b}O)=\mathrm{Tr}_{a}(Af(O)).
\end{equation}

\section{Numerical computation of the partial trace}

\label{num_bip}

In this section we analyze the numerical calculation of the partial
trace for bi-partite systems by first considering the direct implementation
of its definition and afterwards optimizing it by identifying and
avoiding doing null multiplications and sums.

\subsection{Direct implementation}

\label{sec:1st}

Let us analyze the number of basic operations, \emph{scalar multiplications}
(mops) and \emph{scalar sums} (sops), needed to compute the partial
trace directly as given in Eq. (\ref{eq:ptr1}). Considering that
the tensor product of two matrices of dimensions $m\mathrm{x}n$ and
$o\mathrm{x}p$ requires $monp$ mops and zero sops and that the multiplication
of two matrices of dimensions $m\mathrm{x}n$ and $n\mathrm{x}o$
requires $mon$ mops and $mo(n-1)$ sops, we arrive at the numbers
of basic operations shown in Table \ref{table:bo_ptr1}. 
\begin{table}[H]
\begin{center}%
\begin{tabular}{|c|c|c|}
\hline 
Operation & No. of mops & No. of sops\tabularnewline
\hline 
\hline 
$\mathbb{I}_{a}\otimes\langle b_{j}|$ & $d_{a}^{2}d_{b}$ & $0$\tabularnewline
\hline 
$\mathbb{I}_{a}\otimes|b_{j}\rangle$ & $d_{a}^{2}d_{b}$ & $0$\tabularnewline
\hline 
$(\mathbb{I}_{a}\otimes\langle b_{j}|)O$ & $d_{a}^{3}d_{b}^{2}$ & $d_{a}^{2}d_{b}(d_{a}d_{b}-1)$\tabularnewline
\hline 
$(\mathbb{I}_{a}\otimes\langle b_{j}|O)(\mathbb{I}_{a}\otimes|b_{j}\rangle)$ & $d_{a}^{3}d_{b}$ & $d_{a}^{2}(d_{a}d_{b}-1)$\tabularnewline
\hline 
\end{tabular}\end{center}

\caption{Number of basic operations taken by each one of the steps needed to
compute the partial trace when implemented numerically directly from
its definition in Eq. (\ref{eq:ptr1}).}

\label{table:bo_ptr1}
\end{table}
So, to calculate Eq. (\ref{eq:ptr1}) numerically we would make use
of a total of $d_{a}^{2}d_{b}^{2}(2+d_{a}(d_{b}+1))$ mops and $d_{a}^{2}d_{b}(d_{a}d_{b}-1)(d_{b}+1)$
sops. If $d_{b}\gg1$ then
\begin{equation}
\mbox{mops}=\mbox{sops}\approx d_{a}^{3}d_{b}^{3}=d^{3}
\end{equation}
would be needed, where $d=\dim\mathcal{H}$. As the complexity for
the multiplication is, in the ``worst case'', the square of that for
the addition, then this ops is about $\mathcal{O}(d^{6})$.

\subsection{(Not) Using the zeros of $\mathbb{I}_{a}$}

\label{sec:2nd}

Now, instead of simply sending the matrices to a subroutine that computes
tensor products, let's observe that, for $d_{a}\gg1$, most matrix
elements of the identity operator are null. Thus, using the notation:
\begin{equation}
[\alpha,\beta:\gamma]=[O_{\alpha,\beta}\mbox{ }O_{\alpha,\beta+1}\mbox{ }\cdots\mbox{ }O_{\alpha,\gamma-1}\mbox{ }O_{\alpha,\gamma}],
\end{equation}
and with $|0\rangle$ being the null vector in $\mathcal{H}_{b}$,
$p=d_{a}$, and $q=d_{b}$, follows that each term $(\mathbb{I}_{a}\otimes\langle b_{j}|)O(\mathbb{I}_{a}\otimes|b_{j}\rangle)$
in Eq. (\ref{eq:ptr1}) is equal to: 
\begin{eqnarray}
 &  & \mathbb{I}_{a}\otimes\langle b_{j}|\begin{bmatrix}[1,(1-1)q+1:q] & [1,(2-1)q+1:2q] & \cdots & [1,(p-1)q+1:pq]\\{}
[2,(1-1)q+1:q] & [2,(2-1)q+1:2q] & \cdots & [2,(p-1)q+1:pq]\\
\vdots & \vdots & \ddots & \vdots\\{}
[pq,(1-1)q+1:q] & [pq,(2-1)q+1:2q] & \cdots & [pq,(p-1)q+1:pq]
\end{bmatrix}\begin{bmatrix}|b_{j}\rangle & |0\rangle & \cdots & |0\rangle\\
|0\rangle & |b_{j}\rangle & \cdots & |0\rangle\\
\vdots & \vdots & \ddots & \vdots\\
|0\rangle & |0\rangle & \cdots & |b_{j}\rangle
\end{bmatrix}
\end{eqnarray}
\begin{equation}
=\begin{bmatrix}\langle b_{j}| & \langle0| & \cdots & \langle0|\\
\langle0| & \langle b_{j}| & \cdots & \langle0|\\
\vdots & \vdots & \ddots & \vdots\\
\langle0| & \langle0| & \cdots & \langle b_{j}|
\end{bmatrix}\begin{bmatrix}[1,(1-1)q+1:q]|b_{j}\rangle & [1,(2-1)q+1:2q]|b_{j}\rangle & \cdots & [1,(p-1)q+1:pq]|b_{j}\rangle\\{}
[2,(1-1)q+1:q]|b_{j}\rangle & [2,(2-1)q+1:2q]|b_{j}\rangle & \cdots & [2,(p-1)q+1:pq]|b_{j}\rangle\\
\vdots & \vdots & \ddots & \vdots\\{}
[pq,(1-1)q+1:q]|b_{j}\rangle & [pq,(2-1)q+1:2q]|b_{j}\rangle & \cdots & [pq,(p-1)q+1:pq]|b_{j}\rangle
\end{bmatrix}
\end{equation}
\begin{equation}
=\begin{bmatrix}{\displaystyle \sum_{\alpha=1}^{q}}b_{j\alpha}^{*}[(1-1)q+\alpha,(1-1)q+1:q]|b_{j}\rangle & \cdots & {\displaystyle \sum_{\alpha=1}^{q}}b_{j\alpha}^{*}[(1-1)q+\alpha,(p-1)q+1:pq]|b_{j}\rangle\\
{\displaystyle \sum_{\alpha=1}^{q}}b_{j\alpha}^{*}[(2-1)q+\alpha,(1-1)q+1:q]|b_{j}\rangle & \cdots & {\displaystyle \sum_{\alpha=1}^{q}}b_{j\alpha}^{*}[(2-1)q+\alpha,(p-1)q+1:pq]|b_{j}\rangle\\
\vdots & \ddots & \vdots\\
{\displaystyle \sum_{\alpha=1}^{q}}b_{j\alpha}^{*}[(p-1)q+\alpha,(1-1)q+1:q]|b_{j}\rangle & \cdots & {\displaystyle \sum_{\alpha=1}^{q}}b_{j\alpha}^{*}[(p-1)q+\alpha,(p-1)q+1:pq]|b_{j}\rangle
\end{bmatrix}
\end{equation}
\begin{equation}
=\begin{bmatrix}{\displaystyle \sum_{\alpha,\beta=1}^{q}}b_{j\alpha}^{*}O_{(1-1)q+\alpha,(1-1)q+\beta}b_{j\beta} & {\displaystyle \sum_{\alpha,\beta=1}^{q}}b_{j\alpha}^{*}O_{(1-1)q+\alpha,(2-1)q+\beta}b_{j\beta} & \cdots & {\displaystyle \sum_{\alpha,\beta=1}^{q}}b_{j\alpha}^{*}O_{(1-1)q+\alpha,(p-1)q+\beta}b_{j\beta}\\
{\displaystyle \sum_{\alpha,\beta=1}^{q}}b_{j\alpha}^{*}O_{(2-1)q+\alpha,(1-1)q+\beta}b_{j\beta} & {\displaystyle \sum_{\alpha,\beta=1}^{q}}b_{j\alpha}^{*}O_{(2-1)q+\alpha,(2-1)q+\beta}b_{j\beta} & \cdots & {\displaystyle \sum_{\alpha,\beta=1}^{q}}b_{j\alpha}^{*}O_{(2-1)q+\alpha,(p-1)q+\beta}b_{j\beta}\\
\vdots & \vdots & \ddots & \vdots\\
{\displaystyle \sum_{\alpha,\beta=1}^{q}}b_{j\alpha}^{*}O_{(p-1)q+\alpha,(1-1)q+\beta}b_{j\beta} & {\displaystyle \sum_{\alpha,\beta=1}^{q}}b_{j\alpha}^{*}O_{(p-1)q+\alpha,(2-1)q+\beta}b_{j\beta} & \cdots & {\displaystyle \sum_{\alpha,\beta=1}^{q}}b_{j\alpha}^{*}O_{(p-1)q+\alpha,(p-1)q+\beta}b_{j\beta}
\end{bmatrix}.
\end{equation}

So, a generic matrix element of $O^{a}=\mathrm{Tr}_{b}(O)$ shall
take the form:
\begin{equation}
O_{kl}^{a}=\sum_{\alpha,\beta=1}^{d_{b}}O_{(k-1)d_{b}+\alpha,(l-1)d_{b}+\beta}\sum_{j=1}^{d_{b}}b_{j\alpha}^{*}b_{j\beta}.\label{eq:Oa_kl}
\end{equation}
To compute each one of the $d_{a}^{2}$ matrix elements above we need
to do $d_{b}^{2}(d_{b}+1)$ mops and $d_{b}(d_{b}^{2}-1)$ sops. Then,
on the total $d_{a}^{2}d_{b}^{2}(d_{b}+1)$ mops and $d_{a}^{2}d_{b}(d_{b}^{2}-1)$
sops will be necessary. For $d_{b}\gg1$ follows that
\begin{equation}
\mbox{mops}=\mbox{sops}\approx d_{a}^{2}d_{b}^{3}.
\end{equation}
We notice thus a decreasing by a multiplicative factor $d_{a}$ in
ops with relation to the previous direct implementation.

\subsection{(Not) Using the zeros of the computational basis}

\label{sec:3rd}

Now let us recall and verify that \emph{the partial trace is base
independent} and use this fact to diminish considerably the number
of basic operations required for its computation. We regard the following
arbitrary basis for $\mathcal{H}_{b}$: $|j\rangle=\sum_{k=1}^{d_{b}}c_{jk}|b_{k}\rangle$,
with $c_{jk}=\langle j|b_{k}\rangle$. Taking the partial trace in
the basis $|j\rangle$,
\begin{eqnarray}
\mathrm{Tr}_{b}(O) & = & \sum_{j=1}^{d_{b}}(\mathbb{I}_{a}\otimes\langle j|)O(\mathbb{I}_{a}\otimes|j\rangle)=\sum_{j=1}^{d_{b}}(\mathbb{I}_{a}\otimes\sum_{k=1}^{d_{b}}c_{jk}^{*}\langle b_{k}|)O(\mathbb{I}_{a}\otimes\sum_{l=1}^{d_{b}}c_{jl}|b_{l}\rangle)\\
 & = & \sum_{j,k,l=1}^{d_{b}}c_{jk}^{*}c_{jl}(\mathbb{I}_{a}\otimes\langle b_{k}|)O(\mathbb{I}_{a}\otimes|b_{l}\rangle)=\sum_{j,k,l=1}^{d_{b}}\langle j|b_{k}\rangle^{*}\langle j|b_{l}\rangle(\mathbb{I}_{a}\otimes\langle b_{k}|)O(\mathbb{I}_{a}\otimes|b_{l}\rangle)\\
 & = & \sum_{j,k,l=1}^{d_{b}}\langle b_{k}|j\rangle\langle j|b_{l}\rangle(\mathbb{I}_{a}\otimes\langle b_{k}|)O(\mathbb{I}_{a}\otimes|b_{l}\rangle)=\sum_{k,l=1}^{d_{b}}\langle b_{k}|b_{l}\rangle(\mathbb{I}_{a}\otimes\langle b_{k}|)O(\mathbb{I}_{a}\otimes|b_{l}\rangle)\\
 & = & \sum_{k,l=1}^{d_{b}}\delta_{kl}(\mathbb{I}_{a}\otimes\langle b_{k}|)O(\mathbb{I}_{a}\otimes|b_{l}\rangle)=\sum_{k=1}^{d_{b}}(\mathbb{I}_{a}\otimes\langle b_{k}|)O(\mathbb{I}_{a}\otimes|b_{k}\rangle),
\end{eqnarray}
is then seem to be equivalent to compute the partial trace with $|b_{j}\rangle$. 

Hence we shall use the \emph{computational basis}
\begin{equation}
|j\rangle=[\delta_{j1}\mbox{ }\delta_{j2}\mbox{ }\cdots\mbox{ }\delta_{jd_{b}}]^{t},
\end{equation}
(with $\delta_{jk}$ being the Kronecker's delta function) to take
partial traces, avoiding multiplying its $d_{b}-1$ null elements
(for each $|j\rangle$). This is done simply by replacing $|b_{j}\rangle$
by $|j\rangle$ in Eq. (\ref{eq:Oa_kl}) to get\footnote{Another, simpler way to get this result is by applying the definition
in Eq. (\ref{eq:ptr_def2}) to $O$ represented as in Eq. (\ref{eq:Orep}),
but with $|b_{j}\rangle$ being the computational basis $|j\rangle$.
So, utilizing $O^{a}=\mathrm{Tr}_{b}(O)=\sum_{k,l}\sum_{j}\langle k|\otimes\langle j|O|l\rangle\otimes|j\rangle|k\rangle\langle l|=\sum_{k,l}O_{k,l}^{a}|k\rangle\langle l|$
and the fact that only the $((l-1)d_{b}+j)$-th element of $|l\rangle\otimes|j\rangle$
is non-null, we obtain Eq. (\ref{eq:ptre}).}
\begin{eqnarray}
O_{kl}^{a} & = & \sum_{j,\alpha,\beta=1}^{d_{b}}\delta_{j\alpha}\delta_{j\beta}O_{(k-1)d_{b}+\alpha,(l-1)d_{b}+\beta}\\
 & = & \sum_{j=1}^{d_{b}}O_{(k-1)d_{b}+j,(l-1)d_{b}+j}.\label{eq:ptre}
\end{eqnarray}

\begin{figure}[t]
\begin{centering}
\includegraphics[scale=0.5]{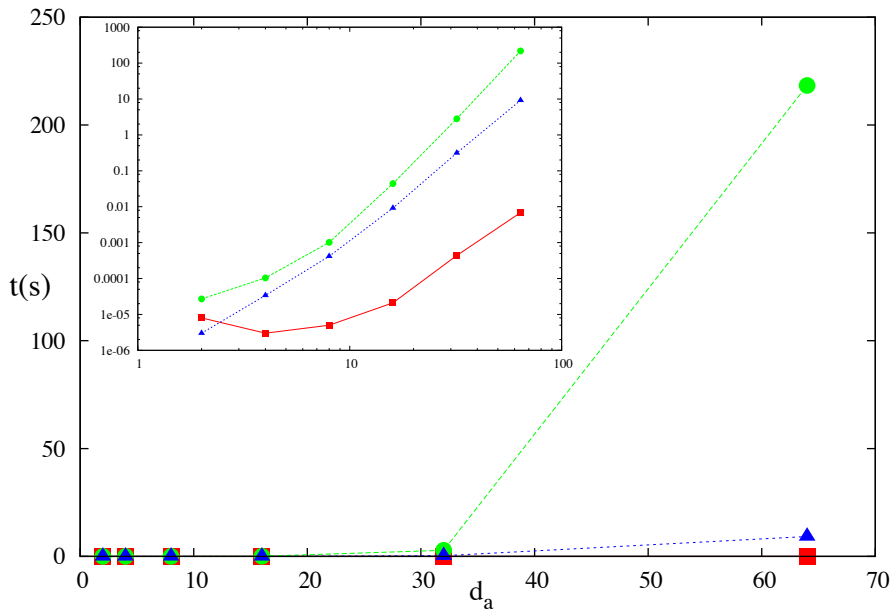}
\par\end{centering}

\caption{(color online) Time taken to compute the partial trace using Eqs.
(\ref{eq:ptr1}), (\ref{eq:Oa_kl}), and (\ref{eq:ptre}) (from up
down) as a function of system $a$ dimension. In the inset is shown
the log-log plot of the same data. We set $d_{b}=d_{a}$ and used
the maximally mixed global state in all cases. The calculations were
performed using the GNU Fortran Compiler version 5.0.0 in a MacBook
Air Processor 1.3 GHz Intel Core i5, with a 4 GB 1600 MHz DDR3 Memory.
If, for $d_{a},d_{b}\gg1$ and $d_{a}=d_{b}$, we have $t\propto d_{a}^{p}d_{b}^{q}$,
then $\log t\propto(p+q)\log d_{a}$. So, the increasing rate of $\log t$
with the number of quantum bits \cite{Nielsen=000026Chuang} constituting
$a$ in the optimized implementation of the partial trace is one fourth
of that for the direct calculation.}

\label{fig:tptr}
\end{figure}

Therefore, in this last implementation of the partial trace operation,
we need to perform ``only'' $d_{a}^{2}(d_{b}-1)$ sops (and no mops).
Or, for $d_{b}\gg1$ 
\begin{equation}
\mbox{mops}=0\mbox{ and }\mbox{sops}\approx d_{a}^{2}d_{b}.
\end{equation}
We think this is the most optimized way to calculate partial traces
for bipartite systems. In the case of Hermitian reduced matrices (in
particular for density matrices) $O_{lk}^{a}=(O_{kl}^{a})^{*}$, and
the number of basic operations needed to compute the partial trace
can be reduced yet by $2^{-1}d_{a}(d_{a}-1)(d_{b}-1)$ sops (which
is almost half of the total when $d_{a},d_{b}\gg1$). For the sake
of illustration, it is shown in Fig. \ref{fig:tptr} the time taken
to compute the partial trace via these three methods as a function
of system $a$ dimension $d_{a}$.

\section{Partial trace for multi-partitions}

\label{PTr_mult}

Let us consider a multipartite system $\mathcal{H}=\bigotimes_{s}\mathcal{H}_{s}$
and a linear operator $O\in\mathcal{L}(\mathcal{H})$. For two arbitrary
sub-systems $s'$ and $s''$, with $s''>s'$, and bases $|s_{j}'\rangle\in\mathcal{H}_{s'}$
and $|s_{j}''\rangle\in\mathcal{H}_{s''}$, one can verify that
\begin{equation}
(\mathbb{I}_{s'}\otimes\mathbb{I}_{(s'+1)\cdots(s''-1)}\otimes|s_{k}''\rangle)(|s_{j}'\rangle\otimes\mathbb{I}_{(s'+1)\cdots(s''-1)})=|s_{j}'\rangle\otimes\mathbb{I}_{(s'+1)\cdots(s''-1)}\otimes|s_{k}''\rangle.
\end{equation}
We then use this relation to see that the equality
\begin{equation}
\mathrm{Tr}_{s's''}(O)=\mathrm{Tr}_{s'}(\mathrm{Tr}_{s''}(O))
\end{equation}
holds for all $s'$ and $s''$. Therefore the partial trace taken
over any set of sub-systems of $\mathcal{H}$ can be implemented sequentially
via partial tracing over single partitions; and we can do that using
the following two main procedures. In one of these procedures we split
$\mathcal{H}$ in two parts $\mathcal{H}_{a}\otimes\mathcal{H}_{b}$
and trace out $a$ or $b$. And in the other one we divide $\mathcal{H}$
in three parties $\mathcal{H}_{a}\otimes\mathcal{H}_{b}\otimes\mathcal{H}_{c}$
and trace over the inner party $b$. As the first procedure was addressed
in the previous section, we shall regard the details of the last one
in this section.

Let $O\in\mathcal{L}(\mathcal{H}_{a}\otimes\mathcal{H}_{b}\otimes\mathcal{H}_{c})$
and let us consider the partial trace
\begin{equation}
O^{ac}=\mathrm{Tr}_{b}(O)=\sum_{j=1}^{d_{b}}(\mathbb{I}_{a}\otimes\langle b_{j}|\otimes\mathbb{I}_{c})O(\mathbb{I}_{a}\otimes|b_{j}\rangle\otimes\mathbb{I}_{c}).
\end{equation}
In a direct implementation of this equation, $d_{a}^{2}d_{b}^{2}d_{c}^{2}(d_{a}d_{c}(d_{b}+1)+2)$
mops and $d_{a}^{2}d_{b}d_{c}^{2}(d_{b}+1)(d_{a}d_{b}d_{c}-1)$ sops
would be used. For $d_{b}\gg1$,
\begin{equation}
\mbox{mops}=\mbox{sops}\approx d_{a}^{3}d_{b}^{3}d_{c}^{3}.
\end{equation}

With the aim of reaching an optimized computation of the partial trace,
we start considering 
\begin{eqnarray}
O^{ac} & = & \sum_{j,m=1}^{d_{a}}\sum_{k,n=1}^{d_{b}}\sum_{l,o=1}^{d_{c}}(\langle j|\otimes\langle k|\otimes\langle l|)O(|m\rangle\otimes|n\rangle\otimes|o\rangle)|j\rangle\langle m|\otimes\mathrm{Tr}_{b}(|k\rangle\langle n|)\otimes|l\rangle\langle o|\\
 & = & \sum_{j,m=1}^{d_{a}}\sum_{l,o=1}^{d_{c}}\left(\sum_{k=1}^{d_{b}}(\langle j|\otimes\langle k|\otimes\langle l|)O(|m\rangle\otimes|k\rangle\otimes|o\rangle)\right)|j\rangle\langle m|\otimes|l\rangle\langle o|\label{eq:ptr_m1}\\
 & = & \sum_{j,m=1}^{d_{a}}\sum_{l,o=1}^{d_{c}}(\langle j|\otimes\langle l|)O^{ac}(|m\rangle\otimes|o\rangle)|j\rangle\langle m|\otimes|l\rangle\langle o|.\label{eq:ptr_m2}
\end{eqnarray}
In this article, if not stated otherwise, \emph{we assume that the
matrix representation of the considered operators in the corresponding
(global) computational basis is given}. Next we notice e.g. that only
the element $\alpha:=(m-1)d_{b}d_{c}+(n-1)d_{c}+o$ of $|m\rangle\otimes|n\rangle\otimes|o\rangle$
is non-null; thus $O|m\rangle\otimes|n\rangle\otimes|o\rangle$ is
equal to the $\alpha$-th column vector of $O$. Thus, Eqs. (\ref{eq:ptr_m1})
and (\ref{eq:ptr_m2}) and these results can be used to write the
following relation between matrix elements (in the corresponding global
computational basis): 
\begin{equation}
O_{(j-1)d_{c}+l,(m-1)d_{c}+o}^{ac}=\sum_{k=1}^{d_{b}}O_{(j-1)d_{b}d_{c}+(k-1)d_{c}+l,(m-1)d_{b}d_{c}+(k-1)d_{c}+o}.
\end{equation}
In terms of ops, in this implementation we shall utilize $\mbox{mops}=0$
and $\mbox{sops}=d_{a}^{2}d_{c}^{2}(d_{b}-1)$, which for $d_{b}\gg1$
is
\begin{equation}
\mbox{mops}=0\mbox{ and }\mbox{sops}\approx d_{a}^{2}d_{b}d_{c}^{2}.
\end{equation}
As in the case of bipartite systems, here also we can utilize the
hermiticity of the reduced matrix to diminish the number of basic
operations by $2^{-1}d_{a}d_{c}(d_{a}d_{c}-1)(d_{b}-1)$ sops.

Fortran code to perform all numerical calculations associated with
this article, and several others, can be accessed in \href{https://github.com/jonasmaziero/LibForQ.git}{https://github.com/jonasmaziero/LibForQ.git}.
In particular, we provide the subroutine, \texttt{partial\_trace(rho,
d, di, nss, ssys, dr, rhor)}, which returns the reduced matrix \texttt{rhor}
once provided \texttt{dr} (its dimension), \texttt{rho }(the matrix
representation in the computational basis of the regarded linear operator
in $\mathcal{H}_{1}\otimes\mathcal{H}_{2}\otimes\cdots\otimes\mathcal{H}_{n}$),
\texttt{d} (the dimension of rho), \texttt{nss}$=n$ (the number of
sub-systems), \texttt{di }(a $n$-dimensional integer vector whose
components specify the dimensions of the sub-systems), and \texttt{ssys
}(a $n$-dimensional integer vector whose null components specify
the sub-systems to be traced over; the other components must be made
equal to one). In the \emph{Hermitian case}, just change the subroutine's
name to \texttt{partial\_trace\_he}.

\begin{figure}
\begin{centering}
\includegraphics[scale=0.65]{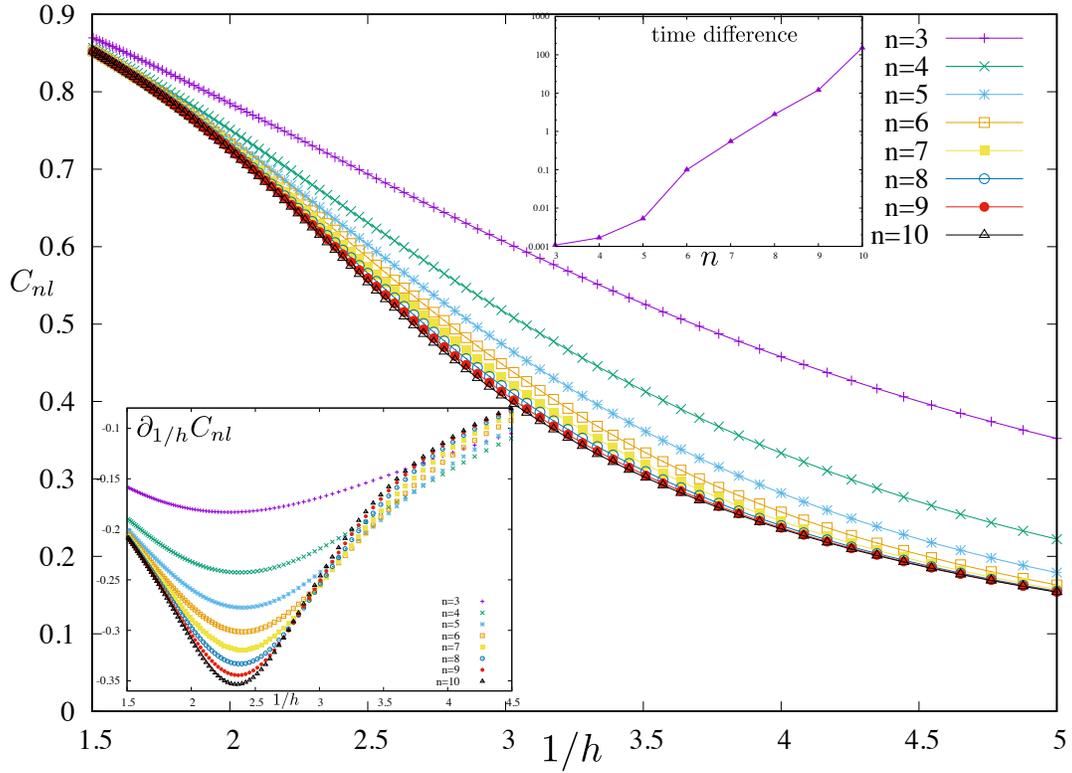}
\par\end{centering}

\caption{(color online) Non-local quantum coherence (NLQC) of the edge qubits
of an Ising chain with $n$ spins as a function of the reciprocal
of the transverse magnetic field. In the bottom inset is shown the
first derivative of the NLQC, whose minimum is seem to go slowly towards
the point of quantum phase transition ($h=1/2$). In the upper inset
is shown, in logarithmic scale, the difference between the time taken,
to do the calculations for a fixed value of $n$, by the direct and
optimized implementations of the partial trace function. We notice
thus a exponential increase of this time difference with $n$. }

\label{fig:inner}
\end{figure}

Let's exemplify the application of what was discussed in this section
by considering the thermal ground state \cite{Nielsen_EntPT}, $\rho=\exp(-\beta H)/\mathrm{Tr}(\exp(-\beta H))$
with $\beta\rightarrow\infty$, of a line of qubits with Ising interaction
between nearest-neighbors: $H=-\frac{J}{2}\sum_{j=1}^{n-1}\sigma_{j}^{z}\sigma_{j+1}^{z}-h\sum_{j=1}^{n}\sigma_{j}^{x},$
where $\sigma_{j}^{x(z)}$ are the Pauli operators in the state space
of the $j$-th spin, $h$ is the so called transverse magnetic field,
and we set the exchange interaction strength to unit ($J=1$). Here
we want to compute the nonlocal quantum coherence \cite{Maziero_QCS}
of the edge spins: $C_{nl}(\rho_{1n})=C(\rho_{1n})-[C(\rho_{1})+C(\rho_{n})],$
where the $l_{1}$-norm quantum coherence is given by \cite{Plenio_QQC}:
$C(\rho)=\sum_{j\ne k}\langle j|\rho|k\rangle,$ with $|j\rangle$
being the standard-computational basis in the regarded Hilbert space.
When performing this kind of calculation, we shall need the reduced
states: $\rho_{1n}=\mathrm{Tr}_{2\cdots(n-1)}(\rho_{12\cdots(n-1)n})\mbox{, }\rho_{1}=\mathrm{Tr}_{n}(\rho_{1n})\mbox{, and }\rho_{n}=\mathrm{Tr}_{1}(\rho_{1n}).$
The results for the quantum coherence and a comparison between the
time taken by the optimized and direct implementations of the partial
trace function are shown in Fig. \ref{fig:inner}.

\section{``Partial traces'' via Bloch's parametrization}

\label{sec:su}

As the most frequent application of the partial trace function is
to compute reduced density matrices (also called partial states or
quantum marginals), let us consider yet another approach we may apply
to perform that task and analyze its computational complexity. From
the defining properties of a density matrix (positiveness $\rho\ge0$
and unit trace $\mathrm{Tr}(\rho)=1$), follows that it can be written
in terms of $\mathbb{I}_{b}$ and of the orthonormal-traceless-hermitian
generators $\Gamma_{j}$ of the special unitary group, as shown below.
For a bipartite system $\mathcal{H}_{a}\otimes\mathcal{H}_{b}$, the
reduced state of sub-system $b$ can be written as \cite{Petruccione_rrho,Bertlmann}:
\begin{equation}
\rho^{b}=\frac{\mathbb{I}_{b}}{d_{b}}+\sum_{j=1}^{d_{b}^{2}-1}\frac{\gamma_{j}}{2}\Gamma_{j},\label{eq:BlochP}
\end{equation}
where $\mathrm{Tr}_{b}(\Gamma_{j}^{\dagger}\Gamma_{k})=2\delta_{jk}$
and $\gamma_{j}=\mathrm{Tr}_{b}(\Gamma_{j}\rho^{b})$. If the density
matrix $\rho$ of the whole system is known, and we want to compute
$\rho^{b}$, then
\begin{equation}
\gamma_{j}=\mathrm{Tr}_{b}(\Gamma_{j}\rho^{b})=\mathrm{Tr}_{ab}(\mathbb{I}_{a}\otimes\Gamma_{j}\rho)=\sum_{k=1}^{d_{a}d_{b}}(\mathbb{I}_{a}\otimes\Gamma_{j}\rho)_{k,k}\label{eq:BlochVC}
\end{equation}
will yield the $d_{b}^{2}-1$ real components of the Bloch vector
$\vec{\gamma}=[\gamma_{1}\mbox{ }\gamma_{2}\mbox{ }\cdots\mbox{ }\gamma_{d_{b}^{2}-1}]^{t}$.

\subsection{Direct implementation}

Let us start by the most straightforward, unoptimized, implementation
of the partial trace via Bloch parametrization. Counting the basic
operations need to compute each $\gamma_{j}$, we note that for the
tensor product $\mathbb{I}_{a}\otimes\Gamma_{j}$, $\mbox{mops}=d_{a}^{2}d_{b}^{2}$,
for the matrix multiplication $(\mathbb{I}_{a}\otimes\Gamma_{j})\rho$,
$\mbox{mops}=d_{a}^{3}d_{b}^{3}$ and $\mbox{sops}=d_{a}^{2}d_{b}^{2}(d_{a}d_{b}-1)$,
and for the trace, $\mbox{sops}=d_{a}d_{b}-1$. Then, for the $d_{b}^{2}-1$
components $\gamma_{j}$ we shall need $\mbox{mops}=(d_{b}^{2}-1)d_{a}^{2}d_{b}^{2}(d_{a}d_{b}+1)$
and $\mbox{sops}=(d_{b}^{2}-1)(d_{a}^{2}d_{b}^{2}+1)(d_{a}d_{b}-1)$.
After knowing $\vec{\gamma}$, $d_{b}^{4}$ mops and $d_{b}^{2}(d_{b}^{2}-2)+d_{b}$
sops are required to compute $\rho^{b}$. Thus, on total we have to
do $\mbox{mops}=d_{b}^{2}(d_{b}^{2}+d_{a}^{2}(d_{b}^{2}-1)(d_{a}d_{b}+1))$
and $\mbox{sops}=(d_{b}^{2}-1)(d_{a}^{2}d_{b}^{2}+1)(d_{a}d_{b}-1)+d_{b}^{2}(d_{b}^{2}-2)+d_{b}$,
which for $d_{b}\gg1$ become
\begin{equation}
\mbox{mops}=\mbox{sops}\approx d_{a}^{3}d_{b}^{5}.
\end{equation}

\subsection{(Not) Using the zeros of \emph{$\mathbb{I}_{a}$}}

Now let's use the zeros of\emph{ $\mathbb{I}_{a}$} to diminish the
number of ops needed to obtain $\gamma_{j}$. We begin by writing
\begin{equation}
(\mathbb{I}_{a}\otimes\Gamma_{j})\rho=\begin{bmatrix}\Gamma_{j} & 0_{d_{b}\mathrm{x}d_{b}} & \cdots & 0_{d_{b}\mathrm{x}d_{b}}\\
0_{d_{b}\mathrm{x}d_{b}} & \Gamma_{j} & \cdots & 0_{d_{b}\mathrm{x}d_{b}}\\
\vdots & \vdots & \ddots & \vdots\\
0_{d_{b}\mathrm{x}d_{b}} & 0_{d_{b}\mathrm{x}d_{b}} & \cdots & \Gamma_{j}
\end{bmatrix}\begin{bmatrix}\rho^{(1)} & [?] & \cdots & [?]\\{}
[?] & \rho^{(2)} & \cdots & [?]\\
\vdots & \vdots & \ddots & \vdots\\{}
[?] & [?] & \cdots & \rho^{(d_{a})}
\end{bmatrix}=\begin{bmatrix}\Gamma_{j}\rho^{(1)} & \Gamma_{j}[?] & \cdots & \Gamma_{j}[?]\\
\Gamma_{j}[?] & \Gamma_{j}\rho^{(2)} & \cdots & \Gamma_{j}[?]\\
\vdots & \vdots & \ddots & \vdots\\
\Gamma_{j}[?] & \Gamma_{j}[?] & \cdots & \Gamma_{j}\rho^{(d_{a})}
\end{bmatrix},
\end{equation}
where $0_{d_{b}\mathrm{x}d_{b}}$ is the $d_{b}\mathrm{x}d_{b}$ null
matrix and $[?]$ is used to denote those $d_{b}\mathrm{x}d_{b}$
sub-blocks of $\rho$ that we do not need when computing $\gamma_{j}$.
From the last equation we see that
\begin{eqnarray}
\gamma_{j} & = & \sum_{\beta=1}^{d_{a}}\mathrm{Tr}(\Gamma_{j}\rho^{(\beta)})=\sum_{\beta=1}^{d_{a}}\sum_{\alpha=1}^{d_{b}}(\Gamma_{j}\rho^{(\beta)})_{\alpha,\alpha}=\sum_{\beta=1}^{d_{a}}\sum_{\alpha=1}^{d_{b}}\sum_{k=1}^{d_{b}}(\Gamma_{j})_{\alpha,k}\rho_{k,\alpha}^{(\beta)}\\
 & = & \sum_{\alpha=1}^{d_{b}}\sum_{k=1}^{d_{b}}(\Gamma_{j})_{\alpha,k}\sum_{\beta=1}^{d_{a}}\rho_{(\beta-1)d_{b}+k,(\beta-1)d_{b}+\alpha},\label{eq:BV_GM}
\end{eqnarray}
which is valid for any choice of the generators $\Gamma_{j}$. If
computed as shown in the last equation, each $\gamma_{j}$ requires
$\mbox{sops}=d_{b}^{2}(d_{a}-1)$ and $\mbox{mops}=d_{b}^{2}$, and
we need to compute $d_{b}^{2}-1$ of them. Thus, when accounted for
also the ops necessary to compute Eq. (\ref{eq:BlochP}) given $\vec{\gamma}$,
we arrive at a total of $2d_{b}^{2}(d_{b}^{2}-1)$ mops and $d_{b}^{2}((d_{b}^{2}-1)d_{a}-1)+d_{b}$
sops needed to compute $\rho^{b}$. For $d_{b}\gg1$:
\begin{equation}
\mbox{mops}\approx2d_{b}^{4}\mbox{ and }\mbox{sops}\approx d_{a}d_{b}^{4}.
\end{equation}

\subsection{(Not) Using the zeros of $\Gamma_{j}$ (and of $\mathbb{I}_{b}$)}

These numbers can be reduced even more if we\emph{ }use the zeros
of the generators $\Gamma_{j}$. To do that, a particular basis $\Gamma_{j}$
has to be chosen, and we shall pick here the generalized Gell Mann's
matrices \cite{Bertlmann}:

\begin{eqnarray}
 &  & \Gamma_{j}^{(1)}=\sqrt{\frac{2}{j(j+1)}}\left(\sum_{k=1}^{j}|k\rangle\langle k|-j|j+1\rangle\langle j+1|\right),\mbox{ for }j=1,\cdots,d_{b}-1,\label{eq:SU1}\\
 &  & \Gamma_{(k,l)}^{(2)}=|k\rangle\langle l|+|l\rangle\langle k|,\mbox{ for }1\leq k<l\leq d_{b},\label{eq:SU2}\\
 &  & \Gamma_{(k,l)}^{(3)}=-i(|k\rangle\langle l|-|l\rangle\langle k|),\mbox{ for }1\leq k<l\leq d_{b}.\label{eq:SU3}
\end{eqnarray}

From Eq. (\ref{eq:BV_GM}), when computing the components of the Bloch
vector, we see that for the generators corresponding to the diagonal
group in Eq. (\ref{eq:SU1}):
\begin{equation}
\gamma_{j}^{(1)}=\sqrt{\frac{2}{j(j+1)}}\sum_{\alpha=1}^{j+1}(-j)^{\delta_{\alpha,j+1}}\sum_{\beta=1}^{d_{a}}\rho_{(\beta-1)d_{b}+\alpha,(\beta-1)d_{b}+\alpha}.
\end{equation}
Then $6(d_{b}-1)$ mops and $1+(d_{a}-1)(\sum_{j=1}^{d_{b}}j)=1+(d_{a}-1)2^{-1}d_{b}(d_{b}+1)$
sops are used for this first group. Related to the generators belonging
to the symmetric and anti-symmetric groups in Eqs. (\ref{eq:SU2})
and (\ref{eq:SU3}), respectively, from Eq. (\ref{eq:BV_GM}) we get
\begin{eqnarray}
\gamma_{(k,l)}^{(2)} & = & \sum_{\beta=1}^{d_{a}}\left(\rho_{(\beta-1)d_{b}+l,(\beta-1)d_{b}+k}+\rho_{(\beta-1)d_{b}+k,(\beta-1)d_{b}+l}\right)=2\sum_{\beta=1}^{d_{a}}\mathrm{Re}(\rho_{(\beta-1)d_{b}+l,(\beta-1)d_{b}+k}),\label{eq:g2}\\
\gamma_{(k,l)}^{(3)} & = & -i\sum_{\beta=1}^{d_{a}}\left(\rho_{(\beta-1)d_{b}+l,(\beta-1)d_{b}+k}-\rho_{(\beta-1)d_{b}+k,(\beta-1)d_{b}+l}\right)=2\sum_{\beta=1}^{d_{a}}\mathrm{Im}(\rho_{(\beta-1)d_{b}+l,(\beta-1)d_{b}+k}).\label{eq:g3}
\end{eqnarray}
These two groups, formed by $d_{b}(d_{b}-1)/2$ elements each, entail
in $\mbox{mops}=2$ and $\mbox{sops}=2(d_{a}-1)$; but, as we will
see below, we do not need them to compute the reduced state $\rho^{b}$.

In the sequence, we shall rewrite the partial state splitting its
diagonal $\Delta$ and non-diagonal $\Theta$ parts, i.e.,
\begin{equation}
\rho^{b}=\Delta+\Theta.
\end{equation}
We will look first at the diagonal elements of $\rho^{b}$:
\begin{eqnarray}
\Delta & = & \frac{\mathbb{I}_{b}}{d_{b}}+\sum_{j=1}^{d_{b}-1}\frac{\gamma_{j}^{(1)}}{2}\Gamma_{j}^{(1)}=\frac{\sum_{k=1}^{d_{b}}|k\rangle\langle k|}{d_{b}}+\sum_{j=1}^{d_{b}-1}\frac{\gamma_{j}^{(1)}}{2}\sqrt{\frac{2}{j(j+1)}}\left(\sum_{k=1}^{j}|k\rangle\langle k|-j|j+1\rangle\langle j+1|\right).
\end{eqnarray}
If, for $j=1,\cdots,d_{b}-1$, we define the function $\xi_{j}=\gamma_{j}^{(1)}/\sqrt{2(j(j+1))}$
and set $\xi_{0}=d_{b}^{-1}$ and $q=d_{b}$, then
\begin{eqnarray}
\Delta & = & \mathrm{diag}(\xi_{0},\xi_{0},\xi_{0},\xi_{0},\xi_{0},\cdots,\xi_{0})+\mathrm{diag}(\xi_{1},-\xi_{1},0,0,0,\cdots,0)+\mathrm{diag}(\xi_{2},\xi_{2},-2\xi_{2},0,0,\cdots,0)\\
 &  & +\mathrm{diag}(\xi_{3},\xi_{3},\xi_{3},-3\xi_{3},0,\cdots,0)+\cdots+\mathrm{diag}(\xi_{q-3},\cdots,\xi_{q-3},-(q-3)\xi_{q-3},0,0)\\
 &  & +\mathrm{diag}(\xi_{q-2},\xi_{q-2},\cdots,\xi_{q-2},-(q-2)\xi_{q-2},0)+\mathrm{diag}(\xi_{q-1},\xi_{q-1},\xi_{q-1},\xi_{q-1},\cdots,\xi_{q-1},-(q-1)\xi_{q-1}).
\end{eqnarray}
After some analysis, we see that
\begin{eqnarray}
\Delta_{1,1} & =\sum_{k=0}^{d_{b}-1}\xi_{k} & \mbox{ and }\Delta_{j,j}=\Delta_{j-1,j-1}+(j-2)\xi_{j-2}-j\xi_{j-1}\mbox{ for }j=2,\cdots,d_{b}.
\end{eqnarray}
If we know the $\gamma_{j}^{(1)}$'s, for the $\xi_{j}$'s $\mbox{mops}=4(d_{b}-1)+1$
and $\mbox{sops}=d_{b}-1$ and for the $\Delta_{j,j}$'s $2\mbox{mops}=\mbox{sops}=4(d_{b}-1)$.
Then for $\Delta$ we shall need $6(d_{b}-1)+1$ mops and $5(d_{b}-1)$
sops.

Let us now consider the off-diagonal elements of $\rho^{b}$:
\begin{eqnarray}
\Theta & = & \sum_{1\le k<l\le d_{b}}\frac{1}{2}\left(\gamma_{(k,l)}^{(2)}\Gamma_{(k,l)}^{(2)}+\gamma_{(k,l)}^{(3)}\Gamma_{(k,l)}^{(3)}\right)=\sum_{1\le k<l\le d_{b}}\frac{1}{2}\left((\gamma_{(k,l)}^{(2)}-i\gamma_{(k,l)}^{(3)})|k\rangle\langle l|+(\gamma_{(k,l)}^{(2)}+i\gamma_{(k,l)}^{(3)})|l\rangle\langle k|\right).
\end{eqnarray}
Using Eqs. (\ref{eq:g2}) and (\ref{eq:g3}) we arrive at
\[
\Theta_{l,k}=\sum_{\beta=1}^{d_{a}}\rho_{(\beta-1)d_{b}+l,(\beta-1)d_{b}+k}.
\]
Then, as $\Theta_{k,l}=\Theta_{l,k}^{*}$, we shall need $\mbox{sops}=2^{-1}d_{b}(d_{b}-1)(d_{a}-1)$
to calculate $\Theta$. So, after knowing $\vec{\gamma}$, $6(d_{b}-1)+1$
mops and $(d_{b}-1)(2^{-1}d_{b}(d_{a}-1)+5)$ sops are used to compute
$\rho^{b}$. Therefore, counting the ops need to compute the $\gamma_{j}^{(1)}$'s,
on total $\mbox{mops}=12(d_{b}-1)+1$ and $\mbox{sops}=d_{b}^{2}(d_{a}-1)+5(d_{b}-1)+1$
shall be used to get $\rho^{b}$ via Bloch parametrization. And for
$d_{a},d_{b}\gg1$
\begin{equation}
\mbox{mops}\approx12d_{b}\mbox{ and }\mbox{sops}\approx d_{a}d_{b}^{2}.
\end{equation}
As the ``worst case'' complexity for the multiplication is equal to
the square of the complexity for the addition, we see that this number
of operations is comparable with the optimized one obtained in Sec.
\ref{sec:3rd}, for $d_{a},d_{b}\gg1$. Here is another, more dramatic
example of the reduction in the number of elementary operations needed
to compute the partial trace ($\mathcal{O}(d_{a}^{6}d_{b}^{10})\rightarrow\mathcal{O}(d_{a}d_{b}^{2})$),
which is obtained simply by performing some analytical developments.
We also provide the Fortran code to compute partial states of bipartite
systems via Bloch's parametrization using this last procedure, i.e.,
with generalized Gell Mann's matrices.

\section{Concluding remarks}

\label{sec:cc}

In this article, a thorough discussion about the partial trace function
was made. We gave special attention to its numerical implementation.
It is a good programming practice trying to avoid making the computer
perform calculations which make no difference to the final result.
We showed here that by following this truism we can decrease considerably
the number of elementary operations needed to compute the partial
trace, or reduced density matrix, which is an extremely important
and common procedure in the quantum mechanics of composite (and open)
systems. We provided and described Fortran code for computing the
partial trace over any set of parties of a discrete multipartite system.
At last, we analyzed the calculation of partial states via the Bloch's
parametrization with generalized Gell Mann's matrices. We believe
this text will be of pedagogical and practical value for the physics
and quantum information science communities. As the partial trace
function is highly adaptable for parallel calculations, we think this
is a natural theme for future investigations. Extending our approach
to continuous variable systems is an interesting research topic. It
would also be fruitful producing translations of our code to other
open source programming languages such as e.g. Maxima and Octave,
as was already done for \href{https://bitbucket.org/huberfe/quantum-code/src/03994b9fc2a1251b82996cea14d65c5e4e2eb216/qgeo.py?fileviewer=file-view-default\#qgeo.py-1456}{Python}.
\begin{acknowledgments}
This work was supported by the Brazilian funding agencies: Conselho Nacional de Desenvolvimento Cient\'ifico e Tecnol\'ogico (CNPq), processes 441875/2014-9 and 303496/2014-2, Instituto Nacional de Ci\^encia e Tecnologia de Informa\c{c}\~ao Qu\^antica (INCT-IQ), process 2008/57856-6, and Coordena\c{c}\~ao de Desenvolvimento de Pessoal de N\'{i}vel Superior (CAPES), process 6531/2014-08. I gratefully acknowledge the hospitality of the Physics Institute and Laser Spectroscopy Group at the Universidad de la Rep\'{u}blica, Uruguay. I also thank Felix Huber for pointing out an error in one of the partial traces subroutines of a previous version of LibForQ.\end{acknowledgments}


\begin{thebibliography}{10}
\bibitem{Feynman82} R. P. Feynman, Simulating physics with computers,
\href{http://link.springer.com/article/10.1007\%2FBF02650179}{Int. J. Theor. Phys. 21, 467 (1982)}.

\bibitem{Loyd} S. Lloyd, Universal quantum simulators, \href{http://science.sciencemag.org/content/273/5278/1073}{Science 273, 1073 (1996)}.

\bibitem{Ito} K. Fisher, H.-G. Matuttis, N. Ito, and M. Ishikawa,
Quantum-statistical simulations for quantum circuits, \href{http://www.worldscientific.com/doi/abs/10.1142/S012918310200367X?journalCode=ijmpc}{Int. J. Mod. Phys. C 13, 931 (2002)}.

\bibitem{Krauth} W. Krauth, Introduction to Monte Carlo algorithms,
\href{http://arxiv.org/abs/cond-mat/9612186}{arXiv:cond-mat/9612186}.

\bibitem{Binder} D. P. Landau and K. Binder, \emph{A Guide to Monte
Carlo Simulations in Statistical Physics} (Cambridge University Press,
New York, 2009).

\bibitem{Kroese} D. P. Kroese, T. Brereton, T. Taimre, and Z. I.
Botev, Why the Monte Carlo method is so important today, \href{http://wires.wiley.com/WileyCDA/WiresArticle/wisId-WICS1314.html}{WIREs Comput. Stat. 6, 386 (2014)}.

\bibitem{Lutsyshyn} Y. Lutsyshyn, Fast quantum Monte Carlo on a GPU,
\href{http://www.sciencedirect.com/science/article/pii/S0010465514003221}{Comp. Phys. Comm. 187, 162 (2015)}.

\bibitem{Kadanoff} L. P. Kadanoff, More is the same; Phase transitions
and mean field theories, \href{http://link.springer.com/article/10.1007/s10955-009-9814-1}{J. Stat. Phys. 137, 777 (2009)}.

\bibitem{Liu_mft} W.-Y. Wang, W.-S. Duan, and J. Liu, The effects
of the beyond mean field corrections of Fermi superfluid gas in a
double-well potential, \href{http://www.worldscientific.com/doi/abs/10.1142/S0129183112500763}{Int. J. Mod. Phys. C 23, 1250076 (2012)}.

\bibitem{Sen} A. Sen(De) and U. Sen, Entanglement mean field theory:
Lipkin\textendash Meshkov\textendash Glick model, \href{http://link.springer.com/article/10.1007\%2Fs11128-011-0279-1}{Quantum Inf. Process. 11, 675 (2012)}.

\bibitem{Yamamoto} D. Yamamoto, G. Marmorini, and I. Danshita, Quantum
phase diagram of the triangular-lattice XXZ model in a magnetic field,
\href{http://journals.aps.org/prl/abstract/10.1103/PhysRevLett.112.127203}{Phys. Rev. Lett. 112, 127203 (2014)}.

\bibitem{Capelle} K. Capelle, A bird\textquoteright s-eye view of
density-functional theory, \href{http://www.scielo.br/scielo.php?script=sci_arttext&pid=S0103-97332006000700035&lng=en&nrm=iso&tlng=en}{Braz. J. Phys. 36, 1318 (2006)}.

\bibitem{Ullrich} C. A. Ullrich and Z.-h. Yang, A brief compendium
of time-dependent density functional theory, \href{http://link.springer.com/article/10.1007\%2Fs13538-013-0141-2}{Braz. J. Phys. 44, 154 (2014)}.

\bibitem{Jones} R. O. Jones, Density functional theory: Its origins,
rise to prominence, and future, \href{http://journals.aps.org/rmp/abstract/10.1103/RevModPhys.87.897}{Rev. Mod. Phys. 87, 897 (2015)}.

\bibitem{Kanjouri_DFT} A. Esmailian, M. Shahrokhi, and F. Kanjouri,
Structural, electronic and magnetic properties of (N, C)-codoped ZnO
nanotube: First principles study, \href{http://www.worldscientific.com/doi/abs/10.1142/S0129183115501302?journalCode=ijmpc}{Int. J. Mod. Phys. C 26, 1550130 (2015)}.

\bibitem{Fisher_RG} M. E. Fisher, Renormalization group theory: Its
basis and formulation in statistical physics, \href{http://journals.aps.org/rmp/abstract/10.1103/RevModPhys.70.653}{Rev. Mod. Phys. 70, 653 (1998)}.

\bibitem{Kadanoff_RG} H. J. Maris and L. P. Kadanoff, Teaching the
renormalization group, \href{http://scitation.aip.org/content/aapt/journal/ajp/46/6/10.1119/1.11224}{Am. J. Phys. 46, 652 (1978)}.

\bibitem{Alvarez} J. V. Alvarez and S. Moukouri, Numerical renormalization
group method in weakly coupled quantum spin chains: Comparison with
exact diagonalization, \href{http://www.worldscientific.com/doi/abs/10.1142/S0129183105007522}{Int. J. Mod. Phys. C 16, 843 (2005)}.

\bibitem{Osborne_RG} C. B\'eny and T. J. Osborne, Information geometric
approach to the renormalisation group, \href{http://journals.aps.org/pra/abstract/10.1103/PhysRevA.92.022330}{Phys. Rev. A 92, 022330 (2015)}.

\bibitem{Orus} R. Oru\'s, A practical introduction to tensor networks:
Matrix product states and projected entangled pair states, \href{http://www.sciencedirect.com/science/article/pii/S0003491614001596}{Ann. Phys. 349, 117 (2014)}.

\bibitem{Cirac} M. Lubasch, J. I. Cirac, and M.-C. Ban\~uls, Algorithms
for finite projected entangled pair states, \href{http://journals.aps.org/prb/abstract/10.1103/PhysRevB.90.064425}{Phys. Rev. B 90, 064425 (2014)}.

\bibitem{Troyer} S. Keller, M. Dolfi, M. Troyer, and M. Reiher, An
efficient matrix product operator representation of the quantum-chemical
Hamiltonian, \href{http://scitation.aip.org/content/aip/journal/jcp/143/24/10.1063/1.4939000}{J. Chem. Phys. 143, 244118 (2015)}.

\bibitem{Karrasch} D. M. Kennesa and C. Karrasch, Extending the range
of real time density matrix renormalization group simulations, \href{http://www.sciencedirect.com/science/article/pii/S0010465515004002}{Comp. Phys. Comm. 200, 37 (2016)}.

\bibitem{Campbell0} S. Campbell, L. Mazzola, and M. Paternostro,
Global quantum correlations in the Ising model, \href{http://www.worldscientific.com/doi/abs/10.1142/S0219749911008404}{Int. J. Quantum Inf. 9, 1685 (2011)}.

\bibitem{Monroe} G.-D. Lin, C. Monroe, and L.-M. Duan, Sharp phase
transitions in a small frustrated network of trapped ion spins, \href{http://journals.aps.org/prl/abstract/10.1103/PhysRevLett.106.230402}{Phys. Rev. Lett. 106, 230402 (2011)}.

\bibitem{Giorgi} G. L. Giorgi and Th. Busch, Genuine correlations
in finite-size spin systems, \href{http://www.worldscientific.com/doi/abs/10.1142/S0217979213450343}{Int. J. Mod. Phys. B 27, 1345034 (2013)}.

\bibitem{Campbell1} S. Campbell, L. Mazzola, G. De Chiara, T. J.
G. Apollaro, F. Plastina, Th. Busch, and M. Paternostro, Global quantum
correlations in finite-size spin chains, \href{http://iopscience.iop.org/article/10.1088/1367-2630/15/4/043033}{New J. Phys. 15, 043033 (2013)}.

\bibitem{Kwek} C. Y. Koh, L. C. Kwek, S. T. Wang, and Y. Q. Chong,
Entanglement and discord in spin glass, \href{http://iopscience.iop.org/article/10.1088/1054-660X/23/2/025202}{Laser Phys. 23, 025202 (2013)}.

\bibitem{Liu} C.-J. Shan, W.-W. Cheng, J.-B. Liu, Y.-S. Cheng, and
T.-K. Liu, Scaling of geometric quantum discord close to a topological
phase transition, \href{http://www.nature.com/articles/srep04473}{Sci. Rep. 4, 4473 (2014)}.

\bibitem{Sen_SwNqb} A. Biswas, R. Prabhu, A. Sen(De), and U. Sen,
Genuine-multipartite-entanglement trends in gapless-to-gapped transitions
of quantum spin systems, \href{http://journals.aps.org/pra/abstract/10.1103/PhysRevA.90.032301}{Phys. Rev. A 90, 032301 (2014)}.

\bibitem{Campbell2} M. J. M. Power, S. Campbell, M. Moreno-Cardoner,
and G. De Chiara, Nonclassicality and criticality in symmetry-protected
magnetic phases, \href{http://journals.aps.org/prb/abstract/10.1103/PhysRevB.91.214411}{Phys. Rev. B 91, 214411 (2015)}.

\selectlanguage{brazil}%
\bibitem{Nielsen=000026Chuang}\foreignlanguage{english}{ M. A. Nielsen
and I. L. Chuang, \emph{Quantum Computation and Quantum Information}
(Cambridge University Press, Cambridge, 2000).}

\selectlanguage{english}%
\bibitem{Preskill} J. Preskill, \emph{Quantum Information and Computation},
\href{http://theory.caltech.edu/people/preskill/ph229/}{http://theory.caltech.edu/people/preskill/ph229/}.

\bibitem{Paris} M. G. A. Paris, The modern tools of quantum mechanics:
A tutorial on quantum states, measurements, and operations, \href{http://link.springer.com/article/10.1140\%2Fepjst\%2Fe2012-01535-1}{Eur. Phys. J. ST 203, 61 (2012)}.

\bibitem{Wilde} M. M. Wilde, \emph{Quantum Information Theory} (Cambridge
University Press, Cambridge, 2013).

\bibitem{Garola} C. Garola and S. Sozzo, The physical interpretation
of partial traces: Two nonstandard views, \href{http://link.springer.com/article/10.1007\%2Fs11232-007-0093-1}{Theor. Math. Phys. 152, 1087 (2007)}.

\bibitem{Fortin} S. Fortin and O. Lombardi, Partial traces in decoherence
and in interpretation: What do reduced states refer to?, \href{http://link.springer.com/article/10.1007\%2Fs10701-014-9791-3}{Found. Phys. 44, 426 (2014)}.

\bibitem{Winter} B. Groisman, S. Popescu, and A. Winter, Quantum,
classical, and total amount of correlations in a quantum state, \href{http://journals.aps.org/pra/abstract/10.1103/PhysRevA.72.032317}{Phys. Rev. A 72, 032317 (2005)}.

\bibitem{Maziero_MI} J. Maziero, Distribution of mutual information
in multipartite states, \href{http://link.springer.com/article/10.1007\%2Fs13538-014-0184-z}{Braz. J. Phys. 44, 194 (2014)}.

\bibitem{Bennett_EE} C. H. Bennett, H. J. Bernstein, S. Popescu,
and B. Schumacher, Concentrating partial entanglement by local operations,
\href{http://journals.aps.org/pra/abstract/10.1103/PhysRevA.53.2046}{Phys. Rev. A 53, 2046 (1996)}.

\bibitem{Okano} A. Hosoya, A. Carlini, and S. Okano, Complementarity
of entanglement and interference, \href{http://www.worldscientific.com/doi/abs/10.1142/S0129183106008716?journalCode=ijmpc}{Int. J. Mod. Phys. C 17, 493 (2006)}.

\bibitem{Fritzche} T. Radtke and S. Fritzsche, Simulation of n-qubit
quantum systems. II. Separability and entanglement, \href{http://www.sciencedirect.com/science/article/pii/S0010465506001445}{Comp. Phys. Comm. 175, 145 (2006)}.

\bibitem{Loss} B. R\"othlisberger, J. Lehmann, and D. Loss, libCreme:
An optimization library for evaluating convex-roof entanglement measures,
\href{http://www.sciencedirect.com/science/article/pii/S0010465511002840}{Comp. Phys. Comm. 183, 155 (2012)}.

\bibitem{Maziero_RevD} L. C. C\'eleri, J. Maziero, R. M. Serra,
Theoretical and experimental aspects of quantum discord and related
measures, \href{http://www.worldscientific.com/doi/abs/10.1142/S0219749911008374?journalCode=ijqi}{Int. J. Quantum Inf. 9, 1837 (2011)}.

\bibitem{Vedral_rd} K. Modi, A. Brodutch, H. Cable, T. Paterek, and
V. Vedral, The classical-quantum boundary for correlations: Discord
and related measures, \href{http://journals.aps.org/rmp/abstract/10.1103/RevModPhys.84.1655}{Rev. Mod. Phys. 84, 1655 (2012)}.

\bibitem{Maziero_Opt} G. H. Aguilar, O. Jim\'enez Far\'ias, J. Maziero,
R. M. Serra, P. H. Souto Ribeiro, and S. P. Walborn, Experimental
estimate of a classicality witness via a single measurement, \href{http://journals.aps.org/prl/abstract/10.1103/PhysRevLett.108.063601}{Phys. Rev. Lett. 108, 063601 (2012)}.

\bibitem{Maziero_rrho} J. Maziero, Random sampling of quantum states:
A survey of methods, \href{http://link.springer.com/article/10.1007\%2Fs13538-015-0367-2}{Braz. J. Phys. 45, 575 (2015)}.

\bibitem{Maziero_Fort} J. Maziero, Fortran code for generating random
probability vectors, unitaries, and quantum states, \href{http://dx.doi.org/10.3389/fict.2016.00004}{Frontiers in ICT 3, 4 (2016)}.

\bibitem{Sacramento} N. Paunkovi\'c, P. D. Sacramento, P. Nogueira,
V. R. Vieira, and V. K. Dugaev, Fidelity between partial states as
signature of quantum phase transitions, \href{http://journals.aps.org/pra/abstract/10.1103/PhysRevA.77.052302}{Phys. Rev. A 77, 052302 (2008)}.

\bibitem{Zeng} J.-Y. Chen, Z. Ji, Z.-X. Liu, Y. Shen, and B. Zeng,
Geometry of reduced density matrices for symmetry-protected topological
phases, \href{http://journals.aps.org/pra/abstract/10.1103/PhysRevA.93.012309}{Phys. Rev. A 93, 012309 (2016)}.

\bibitem{Bellomo2} G. Bellomo, A. Plastino, and A. R. Plastino, Classical
extension of quantum-correlated separable states, \href{http://www.worldscientific.com/doi/abs/10.1142/S021974991550015X}{Int. J. Quantum Inf. 13, 1550015 (2015)}.

\bibitem{Christandl} M. Christandl, B. Doran, S. Kousidis, and M.
Walter, Eigenvalue distributions of reduced density matrices, \href{http://link.springer.com/article/10.1007\%2Fs00220-014-2144-4}{Commun. Math. Phys. 332, 1 (2014)}.

\bibitem{Modi} L. Chen, O. Gittsovich, K. Modi, and M. Piani, Role
of correlations in the two-body-marginal problem, \href{http://journals.aps.org/pra/abstract/10.1103/PhysRevA.90.042314}{Phys. Rev. A 90, 042314 (2014)}.

\bibitem{Viola1} P. D. Johnson and L. Viola, On state versus channel
quantum extension problems: exact results for $U\otimes U\otimes U$
symmetry, \href{http://iopscience.iop.org/article/10.1088/1751-8113/48/3/035307/meta}{J. Phys. A: Math. Theor. 48, 035307 (2015)}.

\bibitem{Viola2} P. D. Johnson and L. Viola, Compatible quantum correlations:
On extension problems for Werner and isotropic states, \href{http://journals.aps.org/pra/abstract/10.1103/PhysRevA.88.032323}{Phys. Rev. A 88, 032323 (2013)}.

\bibitem{Chen1} J. Chen, Z. Ji, D. Kribs, N. L\"utkenhaus, and B.
Zeng, Symmetric extension of two-qubit states, \href{http://journals.aps.org/pra/abstract/10.1103/PhysRevA.90.032318}{Phys. Rev. A 90, 032318 (2014)}.

\bibitem{Bathia} R. Bhatia, Partial traces and entropy inequalities,
\href{http://www.sciencedirect.com/science/article/pii/S0024379503003860}{Linear Algebra Appl. 370, 125 (2003)}.

\bibitem{Petz_SSAeq} P. Hayden, R. Jozsa, D. Petz, and A. Winter,
Structure of states which satisfy strong subadditivity of quantum
entropy with equality, \href{http://link.springer.com/article/10.1007\%2Fs00220-004-1049-z}{Commun. Math. Phys. 246, 359 (2004)}.

\bibitem{Lieb_ESSA} E. A. Carlen and E. H. Lieb, Bounds for entanglement
via an extension of strong subadditivity of entropy, \href{http://link.springer.com/article/10.1007\%2Fs11005-012-0565-6}{Lett. Math. Phys. 101, 1 (2012)}.

\bibitem{Schlosshauer} M. Schlosshauer, Decoherence, the measurement
problem, and interpretations of quantum mechanics, \href{http://journals.aps.org/rmp/abstract/10.1103/RevModPhys.76.1267}{Rev. Mod. Phys. 76, 1267 (2004)}.

\bibitem{Zurek_Dar} W. H. Zurek, Quantum Darwinism, \href{http://www.nature.com/nphys/journal/v5/n3/full/nphys1202.html}{Nat. Phys. 5, 181 (2009)}.

\bibitem{Maziero_Zimmer} J. Maziero and F. M. Zimmer, Genuine multipartite
system-environment correlations in decoherent dynamics, \href{http://journals.aps.org/pra/abstract/10.1103/PhysRevA.86.042121}{Phys. Rev. A 86, 042121 (2012)}.

\bibitem{Fritzsche2} T. Radtke and S. Fritzsche, Simulation of n-qubit
quantum systems. III. Quantum operations, \href{http://www.sciencedirect.com/science/article/pii/S0010465507001804}{Comp. Phys. Comm. 176, 617 (2007)}.

\bibitem{Arfken} G. B. Arfken and H. J. Weber, \emph{Mathematical
Methods for Physicists} (Elsevier, California, 2005).

\bibitem{Watrous} J. Watrous, \emph{Theory of Quantum Information},
\href{https://cs.uwaterloo.ca/~watrous/TQI/}{https://cs.uwaterloo.ca/$\sim$watrous/TQI/}.

\bibitem{Nielsen_EntPT} T. J. Osborne and M. A. Nielsen, Entanglement
in a simple quantum phase transition, \href{http://dx.doi.org/10.1103/PhysRevA.66.032110}{Phys. Rev. A 66, 032110 (2002)}.

\bibitem{Maziero_QCS} M. B. Pozzobom and J. Maziero, Environment-induced
quantum coherence spreading, \href{https://arxiv.org/abs/1605.04746}{arXiv:1605.04746}.

\bibitem{Plenio_QQC} T. Baumgratz, M. Cramer, and M. B. Plenio, Quantifying
coherence, \href{http://dx.doi.org/10.1103/PhysRevLett.113.140401}{Phys. Rev. Lett. 113, 140401 (2014)}.

\bibitem{Petruccione_rrho} E. Br\"uning, H. M\"akel\"a, A. Messina,
and F. Petruccione, Parametrizations of density matrices, \href{http://www.tandfonline.com/doi/abs/10.1080/09500340.2011.632097}{J. Mod. Opt 59, 1 (2012)}.

\bibitem{Bertlmann} R. A. Bertlmann and P. Krammer, Bloch vectors
for qudits, \href{http://iopscience.iop.org/article/10.1088/1751-8113/41/23/235303/meta}{J. Phys. A: Math. Theor. 41, 235303 (2008)}.\end{thebibliography}
\end{document}